\documentstyle[twocolumn,prb,aps,epsfig]{revtex}

\begin{document}

\title{Stationary solutions of the one-dimensional
nonlinear Schr\"odinger equation: I. Case of repulsive nonlinearity}

\author{L. D. Carr$^{1}$\cite{byline}, Charles W. Clark$^{2}$, W. P. Reinhardt$^{1,2,3}$\\}
\address{$^{1}$Department of Physics, University of Washington, Seattle, WA 98195-1560, USA\\}
\address{$^{2}$Electron and Optical Physics Division, National Institute of Standards and Technology, Technology Administration, US Department of Commerce, Gaithersburg, MD 20899, USA\\}
\address{$^{3}$Department of Chemistry, University of Washington, Seattle, WA 98195-1700, USA\\}

\date{\today}

\maketitle
\begin{abstract}
All stationary solutions to the one-dimensional nonlinear Schr\"odinger equation under box and periodic boundary conditions are presented in analytic form.  We consider the case of repulsive nonlinearity; in a companion paper we treat the attractive case.  Our solutions take the form of stationary trains of dark or grey density-notch solitons.  Real stationary states are in one-to-one correspondence with those of the linear Schr\"odinger equation.  Complex stationary states are uniquely nonlinear, nodeless, and symmetry-breaking.  Our solutions apply to many physical contexts, including the Bose-Einstein condensate and optical pulses in fibers.
\end{abstract}



\narrowtext 



\section{Introduction}
\label{sec:intro}
The nonlinear Schr\"odinger equation (NLSE) 
models many phenomena observed in the recently created dilute gas Bose-Einstein condensates (BECs)\cite{dalfovo1}.  In this context it is also referred to as the Gross-Pitaevski equation\cite{gross1,pitaevskii1}.  The NLSE is ubiquitous.  It describes wave propagation phenomena in many systems besides the BEC, including optical pulses in fibers\cite{kivshar1}, helical excitations of a vortex line\cite{hasimoto1}, Bose-condensed photons\cite{ciao1}, and magnetic films~\cite{kalinikos1}. It is one of a few basic equations upon which the modern theory of integrable nonlinear systems has been founded \cite{drazin1,sulem1}.  

Many applications of the NLSE to BECs have dealt with ground-state properties, but there is growing interest in the possibility of generating topological excitations of a condensate, which may well be described by excited-state solutions of the NLSE.  In this article we investigate such solutions, in the case of a one-dimensional NLSE for repulsive nonlinearity, subject either to box or periodic boundary conditions on a finite interval.  This corresponds to a BEC with repulsive atomic pair interactions, which is the case that has received the most experimental interest; a companion paper \cite{carr16} treats the case of attractive interactions.  The stationary excited states that we study here are related to the well-known soliton solutions of the time-dependent NLSE, and when perturbed give rise to soliton-type motions \cite{reinhardt1}.  Recent experiments show that such motions can be induced in BECs by optical engineering of the condensate phase\cite{denschlag1,burger1}.

Box and periodic boundary conditions are as ubiquitous as the NLSE and give physical insight into the solutions to more complicated potentials~\cite{carr19,carr22}.  They model the potentially quasi-one-dimensional regime of a number of present experiments: the atom waveguide~\cite{key1,dekker1}; the prolate harmonic trap in which is formed a cigar-shaped BEC~\cite{andrews1,burger1,ketterle1}; the newly-developed toroidal trap~\cite{close1}; and finally an oblate harmonic trap with a barrier formed in the middle either by a second spin state of the same atom~\cite{matthews1,williams1} or by a laser~\cite{ketterle1}, in which is formed a pancake-shaped BEC with the center removed.  Periodic boundary conditions provide a first model for toroidal geometries; box boundary conditions are a good starting model for cigar-shaped geometries.

The full spectrum of soliton solutions to the NLSE on the infinite line was discovered by Zakharov and Shabat\cite{zakharov1,zakharov2}.  These authors used the Inverse Scattering Transform, a method to which a great deal of mathematical physics literature has been devoted\cite{drazin1}.  To solve the NLSE under these new boundary conditions we have chosen instead to use straightforward, elementary methods, accessible to a broad spectrum of physicists and simpler than the Inverse Scattering Transform.

There have been many recent applications of Zakharov and Shabat's solutions to the BEC\cite{reinhardt1,morgan2}.  Toroidal\cite{rokhsar1,mueller1} and cylindrical or box-shaped\cite{kivshar5,jackson1} geometries have been considered.  Some authors have solved the parabolic potential numerically\cite{kivshar6,kunze1,carr19}.  However, to the best of our knowledge no one to-date has explored, analytically or otherwise, the full spectrum of bounded, \emph{stationary}, multiple-soliton solutions to the NLSE under periodic and box boundary conditions.

\section{The quasi-one-dimensional NLSE}
\label{sec:form}

The nonlinear Schr\"odinger equation which describes the BEC for $T\ll T_c$ is a three-dimensional mean field theory.  We define the BEC to be in the quasi-one-dimensional regime when its transverse dimensions are on the order of its healing length and its longitudinal dimension is much longer than its transverse ones.  In this case the 1D limit of the 3D NLSE is appropriate, rather than a true 1D mean field theory~\cite{kolomeisky1}, as would be the case for a transverse dimension on the order of the atomic interaction length.

The quasi-1D limit of the 3D NLSE is implicitly used in many places in the literature, a recent example being Liu {\it et al.}\cite{liu1}  Discussions which take into account transverse excitations and various other geometries may be found elsewhere\cite{mueller1,perez1}.  We present a brief derivation in which we require that the wavefunction be approximately separable and that its transverse parts be in the ground state.  By requiring the transverse dimensions to be on the order of the healing length we ensure both separability of the wavefunction and transverse stability of the solutions, as we have numerically illustrated in other works~\cite{carr18,carr22}.

We begin with the NLSE that describes a BEC of $N$ atoms of mass $M$, confined in an external potential $V(\vec{r})$:

\begin{eqnarray}
[-\frac{\hbar^{2}}{2M}\nabla^{2}+g\mid\!\psi(\vec{r},t)\!\mid^{2}+V(\vec{r})\,]\,\psi(\vec{r},t) \nonumber \\
= \imath\hbar\partial_{t}\,\psi(\vec{r},t)
\label{eqn:nlse3D}
\end{eqnarray}
where $\mid\!\!\psi(\vec{r},t)\!\!\mid^{2}$ is the single particle density such that $\rho(\vec{r},t)=N\mid\!\!\psi(\vec{r},t)\!\!\mid^{2}$, the coupling constant $g\equiv 4\pi\hbar^{2}aN/M$, and $a$ is the s-wave scattering length for binary collisions between atoms.  The case of repulsive interactions considered here corresponds to $a>0$. 

$V(\vec{r})$ is defined to be a three-dimensional rectangular box of length L and small transverse area $A_{t}$.  In the transverse directions the wavefunction is required to vanish on the surface of the container; in the longitudinal direction we require either box or periodic boundary conditions.  This models the quasi-one-dimensional regime of many BEC experiments, as was mentioned in Sec.~\ref{sec:intro}, as well as ring lasers\cite{agrawal1,arbel1}, helical excitations of a vortex line or ring\cite{hasimoto1}, and many other physical systems for which the 1D NLSE is a good model.

The characteristic length scale over which the condensate density resumes its average value away from a sharp defect or from a perfectly confining wall is the healing length $\xi$:

\begin{equation}
\label{eqn:xidefinition} 
\xi \equiv (8\pi\bar{\rho}\mid\!a\!\mid)^{-1/2}
\end{equation}
$\bar{\rho}\equiv N/(L A_{t})$ is the mean particle density, where $L$ is the longitudinal length of the confining potential and $L_{y}$ and $L_{z}$ are the transverse lengths.  $A_{t}\equiv L_{y}L_{z}$ is the transverse area.  

The BEC is in the quasi-1D regime when $L_y$ and $L_z$ satisfy the following criteria:  $L_{y},L_{z} \lesssim \xi$ and $L_{y},L_{z}\ll L$.  The former ensures that the condensate remains in the ground state in the two transverse dimensions while the latter ensures that longitudinal excitations are much lower in energy than possible transverse excitations.  Under these conditions one may make an adiabatic separation of longitudinal and transverse variables: $\psi(\vec{r},t)=(L A_{t})^{-1/2}f(x)h(y,z)e^{-\imath \mu t/\hbar}$, where $f(x)$ and $h(y,z)$ are dimensionless and the time dependence of a stationary state has been assumed, $\mu$ being the chemical potential.  

This reduces the three-dimensional NLSE (\ref{eqn:nlse3D}) to:

\begin{eqnarray}
[-\frac{\hbar^{2}}{2M}\nabla^{2}+\frac{g\mid\!f(x)h(y,z)\!\mid^{2}}{L A_{t}}+V(\vec{r})\,]\,f(x)h(y,z) \nonumber \\
= \mu f(x)h(y,z)
\label{eqn:sepVar}
\end{eqnarray}

Eq.~(\ref{eqn:sepVar}) may be projected onto the ground state of $h(y,z)$ and integrated over the transverse dimensions $y,z$:

\begin{eqnarray}
\int_{0}^{L_{y}} dy\int_{0}^{L_{z}} dz\:h_{gs}^{*}(y,z)\,[\mu - \frac{\hbar^{2}}{2M}\nabla^{2}+\frac{g\mid\!f(x)h(y,z)\!\mid^{2}}{L A_{t}} \nonumber \\
+V(x,y,z)\,]\,f(x)h(y,z)=0
\label{eqn:projectNlse}
\end{eqnarray}

As will be shown in Sec.~\ref{subsec:particle}, in the limit that $L_{y},L_{z}\sim\xi$, $h(y,z)$ takes the form of the ground state linear quantum mechanics particle-in-a-box solution $h_{gs}(y,z)=h_{o}sin(\pi y/L_{y})sin(\pi z/L_{z})$.  Requiring $h_{gs}(y,z)$ to be normalized to 1, $h_{o}=2$.  We multiply Eq.~\ref{eqn:projectNlse} through by $2M\xi^2/\hbar^2$ and obtain a simple, quasi-1D NLSE:

\begin{eqnarray}
[-(\tilde{\mu}-\frac{\pi^{2}\xi^{2}}{L_{y}^{2}}-\frac{\pi^{2}\xi^{2}}{L_{z}^{2}})-\frac{\xi^2}{L^2}\partial_{x}^{2}+\frac{9}{4}\mid\!f(x)\!\mid^{2} \nonumber \\
+\tilde{V}(x,y,z)]\,f(x)=0
\end{eqnarray}
where $f$ is the dimensionless wavefunction describing excitations along $L$; $\mid\!f\!\mid^2/L$ is the longitudinal part of the single particle density; $\tilde{V}(\tilde{x})\equiv(2M\xi^{2}/\hbar^{2})V(\tilde{x})$ is the confining potential; and $\tilde{\mu}\equiv(2M\xi^{2}/\hbar^{2})\mu$ is a dimensionless chemical potential which is now the eigenvalue of the problem.  

The notation is further simplified by combining the longitudinal length of the confining potential and the healing length into a single dimensionless scaling parameter:

\begin{equation}
\label{eqn:lambdadef}
\lambda \equiv \xi/L
\end{equation}
$\lambda$ is an important parameter which will determine many of the properties of the stationary states.  For the purpose of mathematical ease, $\tilde{x}\equiv x/L$.  Using the approximations for $L_{y},L_{z}$ and dividing through by the integration factor of $9/4$ results in the dimensionless, 1D NLSE we shall use henceforth:

\begin{equation}
[-\lambda_{eff}^{2}\partial_{\tilde{x}}^{2}+\mid\!f(\tilde{x})\!\mid^{2}+\tilde{V}(\tilde{x})\,]\,f(\tilde{x}) = \tilde{\mu}_{eff}\,f(\tilde{x})
\label{eqn:nlse}
\end{equation}
where $\tilde{\mu}_{eff}=\tilde{\mu}-8\pi^{2}/9$ and $\lambda^{2}_{eff}=4\lambda^{2}/9$.  For the purposes of this presentation $\xi^{2}=4/(9*8\pi\mid\!a\!\mid\bar{\rho})$ and $\tilde{\mu}=(2M\xi^{2}/\hbar^{2})\mu-8\pi^{2}/9$.  However, we shall simply drop the \emph{eff} subscripts, as such constant factors make no difference in our results.

For comparison with experiment the conversion factors from the dimensionless $\tilde{\mu}$ to $\mu$ in $\mu$K are listed.  The general conversion is $\mu=(8.34*10^{-15})(\bar{\rho}\,a/M)\tilde{\mu}$, where $M$ is in atomic mass units, $\bar{\rho}$ is in $\text{cm}^{-3}$, and $a$ is in nm.  Using common experimental values\cite{dalfovo1} of $\bar{\rho} \sim 10^{14}$, for $^{23}$Na $a\sim 2.75$, and for $^{87}$Rb $a\sim 5.77$, the conversion factors are 0.0723 and 0.0401, respectively.  Since the dimensionless chemical potentials found will be on the order of 1 to 10 this gives a sense of the energy scale of the solutions, on the order of 0.1 to 1 $\mu$K.  Note that throughout the presentation an experimentally reasonable test scale of $\lambda=1/25$ will be used for illustrative purposes.

As $\mid\!f(\tilde{x})\!\mid^2$ is a single particle density, it is normalized to 1 rather than $N$:

\begin{equation}
\int_{0}^{1} d\tilde{x} \mid\!f(\tilde{x})\!\mid^2 = 1
\label{eqn:normCond}
\end{equation}
The number of atoms $N$, which is proportional to the coefficient to the nonlinear term in Eq.~(\ref{eqn:nlse3D}), is then contained in the ratio of the healing length to the box length, $\lambda\propto N^{-1/2}$.

The NLSE (\ref{eqn:nlse}), subject to normalization (\ref{eqn:normCond}) and such boundary conditions as will be described below, is the equation we will solve.

\section{Box Boundary Conditions}
\label{sec:bbc}

We now consider the solution of Eq.~(\ref{eqn:nlse}) in 
regions of constant potential, which may be taken to be 
$\tilde{V}(\tilde{x}) = 0$ without loss of generality.  
We note first that if $f(\tilde{x})$
vanishes anywhere in an interval, as for 
example at the edges of the box, 
then $f(\tilde{x})$
may be taken to be purely real throughout that interval.
This is easily established by considering a Taylor series
expansion of $f$ in the neighborhood of the point at
which it vanishes.  Thus, we may remove the absolute
value symbol in Eq.~(\ref{eqn:nlse}) and so recover
an ordinary nonlinear equation for a real function:

\begin{equation}
-\lambda^{2}f^{\prime\prime}+f^{3}-\tilde{\mu}f=0
\end{equation}

By multiplying through by $f$, integrating both sides, and then solving for $d\tilde{x}$, the solution may be written in the form:

\begin{equation}
\tilde{x}=\frac{\sqrt{2}\lambda}{\sqrt{R_{+}}}\int_{0}^{f/\sqrt{R_{-}}} \frac{dt}{\sqrt{1-t^{2}}\sqrt{1-m\,t^{2}}}
\label{eqn:jeSoln}
\end{equation}
where $R_{\pm}\equiv 1\pm\sqrt{1-C}$, with $C$ a constant of 
integration, and $m=R_{-}/R_{+}$.  Comparing Eq.~(\ref{eqn:jeSoln}) 
to Eq.~(\ref{eqn:snDefn}) in App.~\ref{app:jacobian}, 
it is apparent that they differ only by trivial scaling factors.  
Therefore in the box the most general solution 
is a Jacobian elliptic function, which as shown in 
App.~\ref{app:jacobian} must be the sn function.  A brief review of the form and properties of the Jacobian elliptic functions is given in the appendix.

\subsection{Solutions and spectra}

The most general form of the solution is:

\begin{equation}
f(\tilde{x})=A\,\text{sn}(k \tilde{x}+\delta\mid m)
\label{eqn:boxAnsatz}
\end{equation}
where the notation sn$(x\mid m)$ is standard, 
as used in App.~\ref{app:jacobian}.  
$k$ and $\delta$ will be determined 
by the boundary conditions below while 
$A$ and $m$ will be determined by 
substitution of Eq.~(\ref{eqn:boxAnsatz}) 
into the NLSE and by normalization.

The boundary conditions are:

\begin{equation}
f(0)=f(1)=0
\label{eqn:boxBC}
\end{equation}
The boundary condition at the origin can be
satisfied most easliy by taking $\delta = 0$.
The function sn$(x\mid m)$ is periodic in
x, with period equal to  $4K(m)$, where $K(m)$
is an elliptic integral of the first kind (see App. A). 
Thus the boundary equations at $\tilde{x}=0$ and $\tilde{x}=1$ are 
satisfied if $k=2jK(m)$, where $j\in\{1,2,3,...\}$.  
The  number of 
nodes in the jth solution is 
$j-1$.  We will give a more general interpretation of
 $j$ below.  We then solve Eq.~(\ref{eqn:nlse}) by 
substituting Eq.~(\ref{eqn:boxAnsatz}), 
using Jacobian elliptic identities, and 
setting coefficients of equal powers of sn equal.  
This results in equations for the amplitude, $A$,
and the chemical potential, $\tilde{\mu}$:

\begin{equation}
A^{2}=2 m (2 j K(m))^{2}\lambda^2
\label{eqn:bbcAmp}
\end{equation}

\begin{equation}
\tilde{\mu}= (2 j K(m))^{2}\lambda^2(1+m)
\label{eqn:bbcEnergy}
\end{equation}

Substituting Eq.~(\ref{eqn:bbcAmp}) into Eq.~(\ref{eqn:normCond}), and noting that the integral over sn$^{2}$ can be defined in multiples of the quarter period $K(m)$, we obtain the normalization condition:

\begin{equation}
2 (2 j K(m))^{2}\lambda^2(1-\frac{E(m)}{K(m)})=1
\label{eqn:bbcNormCond}
\end{equation}
where $E(m)$ is the complete elliptic integral of the second kind.  Eq.~(\ref{eqn:boxAnsatz}) then becomes:

\begin{equation}
f(\tilde{x})=\sqrt{2 m}(2 j K(m))\lambda\,\text{sn}(2 j K(m) \tilde{x}\mid m)
\label{eqn:bbcWavefn}
\end{equation}

\begin{figure}
\begin{center}
\epsfig{figure=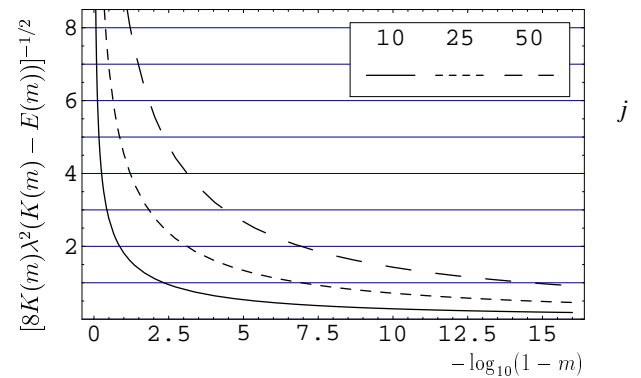,width=8.2cm}
\end{center}
\caption{
This graphical solution of Eq.~(\ref{eqn:bbcNormCond}) shows that for a given scale and number of nodes the real solution to the stationary NLSE under box or periodic boundary conditions is unique.  $\lambda$ is the scale and $j-1$ with $j\in\{1,2,3,...\}$ or $j$ with $j\in\{2,4,6,...\}$ is the number of nodes, respectively.  The three curved lines are plots of Eq.~(\ref{eqn:bbcNormCond}) solved for the number of nodes $j$, with $\lambda^{-1}=L/\xi=10,25,50$.  The left-hand side of the plot is the $m=0$, linear limit, while the right-hand side exponentially approaches the $m=1$, topological soliton limit.  The solutions are found where these lines intersect with the horizontal lines of $j$.  Note the rapid convergence to $m=0$ in the high $j$ limit, so that for large $j$ the solutions are in the linear regime.
}
\label{fig:FnormCond}
\end{figure}

This leaves the chemical potential (\ref{eqn:bbcEnergy}) and the wavefunction (\ref{eqn:bbcWavefn}) determined up to the parameter $m$ and the scale $\lambda$.  In Fig.~\ref{fig:FnormCond} a graphical solution of Eq.~(\ref{eqn:bbcNormCond}) is shown.  The plot demonstrates that the solutions are unique.  Such solutions are in one-to-one correspondence with those of the 1D particle-in-a-box problem in linear quantum mechanics.
 
\begin{figure}
\begin{center}
\epsfig{figure=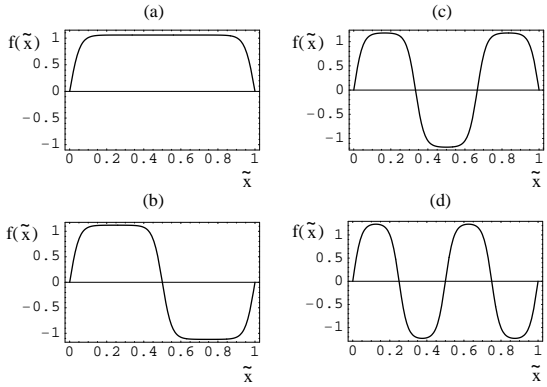,width=8.2cm}
\end{center}
\caption{
Real stationary solutions to the NLSE under box and periodic boundary conditions.  These are in one-to-one correspondence with those of the analogous particle-in-a-box and particle-on-a-ring problems in linear quantum mechanics, and may also be characterized as dark soliton trains.  Box: (a)-(d) are the ground state and first three excited states.  Ring: (b) and (d) are the first two solutions of this type.  Chemical potentials: (a) $\tilde{\mu}=1.120$ (b) $\tilde{\mu}=1.253$ (c) $\tilde{\mu}=1.402$ (d) $\tilde{\mu}=3.028$.  All plots are for the test scale of $\xi/L=1/25$.
}
\label{fig:FbbcProbAmp}
\end{figure}

Plots of the wavefunction for the ground state and the first three excited states are shown in Fig.~\ref{fig:FbbcProbAmp}.  To meet the box boundary conditions the wavefunction drops to zero over the scale of the healing length $\xi$.  When the zeros of $f$ are well-separated, the analytic behaviour of $f$ near a zero, $\tilde{x}_o$, approaches $f\sim\text{tanh}((\tilde{x}-\tilde{x}_o)/(\lambda\sqrt{2}))=\text{tanh}((x-x_o)/(\xi\sqrt{2}))$.  We refer to this behaviour at each node of $f$ as a kink.  $f^2$ is proportional to the density of particles in a BEC; this density is constant everywhere except at the boundaries and the kinks, where it dips to zero.  

\begin{figure}
\begin{center}
\epsfig{figure=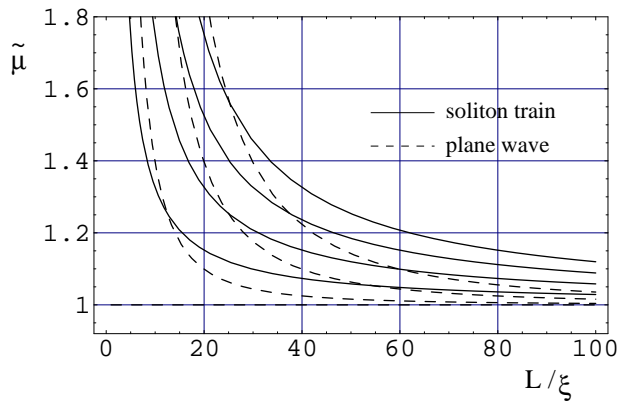,width=8.2cm}
\end{center}
\caption{
Chemical potential spectra of real stationary states, as a function of inverse scale $L/\xi$, with stationary plane wave spectra shown for comparison.  Solid lines: shown are $n=0,1,2,3$, where $n$ is the phase quantum-number of the plane wave on the ring.  Dashed lines: real stationary states of the NLSE in a box and on a ring are soliton-trains.  Shown are $j=1,2,3,4$ with $j-1$ the number of nodes in a box and $j=2,4$ the number of nodes on a ring.  Note that for very fine scale, i.e. $L/\xi$ large, the chemical potentials are evenly spaced.  This corresponds to the topological soliton limit in which the chemical potentials are additive, just as for vortices.
}
\label{fig:FbbcSpectrum}
\end{figure}

In Fig.~\ref{fig:FbbcSpectrum} we plot the chemical potential spectrum of this solution type as a function of $\lambda^{-1}$, the number of healing lengths per box length.  The leftmost portion of the plot corresponds to the particle-in-a-box limit and the rightmost portion to the topological soliton limit.  We now discuss these two limits.

\subsection{Particle-in-a-box limit}
\label{subsec:particle}

High chemical potential states in which the kinks overlap become particle-in-a-box type solutions, as can be seen in Fig.~\ref{fig:FbbcProbAmp}d.  This is both the zero density, linear limit and the highly-excited-state limit.  Mathematically, $m\rightarrow 0^{+}$ and $\text{sn}\rightarrow \sin$.  Physically, $j\lambda\gg 1$.  In this limit $K(m)\rightarrow \pi(1/2 + m/8 + {\cal O}(m^2))$ and $m\rightarrow 1/(j\pi\lambda)^2$, so that Eq.~(\ref{eqn:bbcEnergy}) becomes:

\begin{eqnarray}
\tilde{\mu}= j^2\pi^2\lambda^2(1+\frac{3m}{2}+{\cal O}(m^2)) \nonumber \\
\tilde{\mu}= j^2\pi^2\lambda^2(1 +\frac{3}{2j^2\pi^2\lambda^2}+{\cal O}(\frac{1}{j^4\lambda^4}))
\end{eqnarray}
corresponding to:

\begin{eqnarray}
\label{eqn:perturb}
\mu= \frac{j^{2}\pi^{2}\hbar^{2}}{2 M L^{2}}(1+\frac{3m}{2}+{\cal O}(m^2)) \nonumber \\
\mu= \frac{j^{2}\pi^{2}\hbar^{2}}{2 M L^{2}}(1 +\frac{12 a N L}{A_t j^2\pi}+{\cal O}(\frac{L^2 N^2}{j^4}))
\end{eqnarray}
which clearly converges to the well-known linear quantum mechanics particle-in-a-box chemical potential as $m\rightarrow 0^{+}$.

One may also obtain this result from first order perturbation theory.  The Hamiltonian for Eq.~(\ref{eqn:nlse}) is $H=H_o + H_1$, where $H_o=-(\hbar^2/2M)\partial_{x}^{2}$, $H_1=(g/(LA_t))f^2$, and the box boundary conditions are implicit.  As the solutions are real, we have dropped the absolute value sign in $H_1$.  Note that we have put the units back in.  The solution to $H_o f = \mu^{(0)} f$ is $f=\sqrt{2}\sin (\pi j \tilde{x})$ with $\mu^{(0)}=(j^{2}\pi^{2}\hbar^{2})/(2 M L^2)$.  The first order perturbation-correction to $\mu^{(0)}$ yields:

\begin{equation}
\mu^{(1)}=\frac{4\pi\hbar^2aN}{M}\frac{1}{LA_t}\frac{3}{2}
\end{equation}
where we have substituted in the definition of the coupling constant $g$.  By noting that $A^2=2$ in Eq.~(\ref{eqn:bbcAmp}) in this limit, and using Eq.~(\ref{eqn:bbcAmp}) together with Eq.~(\ref{eqn:xidefinition}) to eliminate $Na$ in favor of the parameter $m$, one recovers the same first order perturbation-correction as in Eq.~(\ref{eqn:perturb}):

\begin{equation}
\mu^{(1)}=\frac{j^{2}\pi^{2}\hbar^{2}}{2 M L^{2}}\frac{3m}{2}
\end{equation}

\subsection{Topological soliton limit}

One may add a kink without disturbing another kink, provided that the overlap between them is exponentially small in the ratio of their separation to the healing length.  In analogy with vortices the chemical potentials of the kinks ought to be additive.  This is the large particle number, highly nonlinear, Thomas-Fermi\cite{dalfovo1} limit.  Mathematically, $m\rightarrow 1^{-}$ and sn $\rightarrow$ tanh, formally called a topological soliton.  Physically, $N\rightarrow\infty$ implies $\lambda \rightarrow 0^{+}$, so that we expect the kinks should never overlap.  We note that in a BEC experiment the box length is held fixed while atoms condense.

By solving for $K(m)$ in Eq.~(\ref{eqn:bbcNormCond}) and using $E(m)\rightarrow 1^+$ (see App.~\ref{app:jacobian}) we find that $K(m)\rightarrow\kappa$ where:

\begin{equation}
\kappa=\frac{1}{2}(1+\sqrt{1+\frac{1}{2j^{2}\lambda^2}})
\end{equation}
so that Eq.~(\ref{eqn:bbcEnergy}) becomes:

\begin{equation}
\tilde{\mu}= 2(2j)^{2}\lambda^2\kappa^{2}
\label{eqn:solitonEnergy}
\end{equation}
while Eq.~(\ref{eqn:bbcWavefn}) becomes:

\begin{equation}
f(\tilde{x})=\sqrt{2}(2j\kappa)\lambda\,\text{sn}(2j\kappa \tilde{x}\mid m)
\end{equation}

We have found that this limit suffices to calculate chemical potentials for which $j<(1/(5\lambda))$ to better than $1\%$.  This estimate assumes an overall scale size of $\sim 5\xi$ per kink.  The chemical potentials for the $j=1$, $j=2$, and $j=3$ solutions shown in Fig.~\ref{fig:FbbcProbAmp} satisfy this criterion, for example, as do any ground states for a healing length of smaller than $1/10$.  If we now further require that $(8j^{2}\lambda^2) \ll 1$ 
then Eq.~(\ref{eqn:solitonEnergy}) becomes:

\begin{equation}
\tilde{\mu}= 1+2\sqrt{2}j\lambda
\label{eqn:solitonEnergy2}
\end{equation}
from which we see that the chemical potentials of additional kinks are indeed additive.  This additivity is apparent in Fig.~\ref{fig:FbbcSpectrum} in the limit $\lambda^{-1}\equiv L/\xi\rightarrow\infty$.  Eq.~(\ref{eqn:solitonEnergy2}) is identical in form to that of the harmonic oscillator in linear quantum mechanics.

Putting back in the units of $\hbar^2/(2M\xi^2)$, we find that 
the chemical potential for formation of
an additional kink, $\Delta\mu$, is proportional to $\sqrt{N}$ :

\begin{equation}
\Delta\mu= 2\sqrt{2}j\frac{\hbar^2}{2M}\sqrt{\frac{8\pi N a}{ A_t}}\frac{1}{L^{3/2}}
\end{equation}
so that the chemical potential to add a kink increases as atoms condense, when all units are included.

We may also solve for the excitation energy to add an isolated kink to the $N$ body system by finding the expectation value of the many-body Hamiltonian\cite{gross1}:

\begin{equation}
\langle\tilde{H}\rangle = N [\:\lambda^2 \int \mid\!f^{\prime}\!\mid^2 - \frac{1}{2}\int \mid\!f\!\mid^4\:]
\end{equation}
where just as for the chemical potential, $\langle\tilde{H}\rangle \equiv (2M\xi^2/\hbar^2)\langle H\rangle$ is a dimensionless energy.  Substituting in the stationary solution Eq.~(\ref{eqn:boxAnsatz}), one finds:

\begin{equation}
\langle\tilde{H}\rangle= N \frac{2^5}{3}j^4 K(m)^3 \lambda^4 [\:(1+2m)K(m)-(1+m)E(m)\:]
\end{equation}
which in the limit as $(8j^{2}\lambda^2) \ll 1$ becomes:

\begin{equation}
\langle\tilde{H}\rangle= N (\frac{1}{2}+2\sqrt{2}j\lambda+{\cal O}(j^2\lambda^2))
\end{equation}

Therefore the excitation energy to add a kink is $\Delta \langle\tilde{H}\rangle=2N\sqrt{2}\lambda$, or putting the units back in:

\begin{equation}
\Delta\langle H\rangle = \frac{\hbar^2 2^3\sqrt{\pi a}A_t}{2M} \bar{\rho}^{3/2}
\end{equation}
This is just as one would expect; the energy to add a kink no longer depends on the box length when the kink size is much smaller than the box.  Instead it simply depends on the density.

\section{Periodic boundary conditions}
\label{sec:pbc}

There are three solution types for periodic boundary conditions.  There are constant amplitude solutions which are plane waves; real symmetry-breaking solutions, similiar to those found in Sec.~\ref{sec:bbc}; and a novel class of complex symmetry-breaking solutions.  The former two are in one-to-one correspondence with particle-on-a-ring solutions in linear quantum mechanics, while the latter one is only found in the presence of nonlinearity.  As the ring is rotationally invariant, the symmetry-breaking solutions will have a high degeneracy, in analogy with vortices in two dimensions\cite{kosterlitz1}.  The periodic boundary conditions are:

\begin{equation}
f(0)=f(1)
\label{eqn:pbcRealBC1}
\end{equation}

\begin{equation}
f^{\prime}(0)=f^{\prime}(1)
\label{eqn:pbcRealBC2}
\end{equation}

\subsection{Constant amplitude solutions}
\label{subsec:cas}

If we assume that the amplitude is constant then we obtain plane wave solutions of the form:

\begin{equation}
f(\tilde{x})=e^{\imath 2\pi n \tilde{x}}
\label{eqn:constAmp}
\end{equation}
where $n \in \{0,\pm 1,\pm 2, ...\}$.  The amplitude is constrained by normalization to be 1.  Substituting Eq.~(\ref{eqn:constAmp}) into Eq.~(\ref{eqn:nlse}) we find the chemical potential:

\begin{equation}
\tilde{\mu}=1+(\:\lambda2\pi n\:)^{2}
\end{equation}
from which we obtain the lower limit of the chemical potentials under periodic boundary conditions, $\tilde{\mu}=1$.  This is just what we expect physically for the repulsive BEC.  The ground state on a ring is the condensate spread out evenly.  There is no symmetry breaking.  For $n\neq 0$ each solution is two-fold degenerate, as $n$ can be either positive or negative, while the $n=0$, ground state solution is non-degenerate.

Note that these states could also be termed angular momentum eigenstates or quantized vortices, as for example in the work of Matthews {\it et al.}\cite{matthews1}.

\subsection{Real symmetry-breaking solutions}
\label{subsec:pbcreal}

As we have exchanged the ring for the box, Eq.~(\ref{eqn:boxAnsatz}) is the real solution.  One simply changes $k$ from $2jK(m)$ to $4jK(m)$ in order to satisfy Eqs.(\ref{eqn:pbcRealBC1}) and (\ref{eqn:pbcRealBC2}), i.e. from multiples of the half period to multiples of the whole period.  The number of nodes will be 2j rather than j-1, where $j\in\{1,2,3,...\}$.  We temporarily keep $\delta$ set to 0.  But note that, unlike for box boundary conditions, under periodic boundary conditions $\delta$ is arbitrary.  

Then all the results from section~\ref{sec:bbc} hold with the new $k$, by letting $j\rightarrow 2j$ in all equations.  The energy and wavefunction are determined uniquely by graphical solution of Fig.~\ref{fig:FnormCond}.  In Fig.~\ref{fig:FbbcProbAmp} we show the first two states.  Both the linear quantum mechanics, particle-on-a-ring limit and the topological soliton limits are reproduced.  In the latter the same kind of non-overlapping criterion applies as before.  Thus given $(16j^{2}\lambda^{2}) \ll 1$ Eq.~(\ref{eqn:bbcEnergy}) becomes:

\begin{equation}
\tilde{\mu}= 1+2\sqrt{2}(2j)\lambda
\end{equation}
Note that the factor of two in front of $j$ shows that, on a ring, kinks of this type come in pairs.

If $\delta$ is permitted to vary arbitrarily, the degeneracy inherent in these symmetry-breaking solutions is obtained.  The entropy associated with $j$ kinks depends logarithmically on the box length $L$, and, since there are approximately $\lambda^{-1}$ possible positions for the kink, the entropy is:

\begin{equation}
S\sim k_{\text{B}}ln(\frac{1}{4\sqrt{2}j\lambda})
\label{eqn:entropy1}
\end{equation}
where the factor of $4\sqrt{2}$ comes from $2\sqrt{2}\,\xi$ for each of the two kinks.  This is consistent with the non-overlapping criterion we used in obtaining Eq.~(\ref{eqn:solitonEnergy2}).

\subsection{Complex symmetry-breaking solutions}
\label{subsec:pbccomplex}

For complex solutions we divide the wavefunction into a phase and amplitude:

\begin{equation}
f(\tilde{x})=r(\tilde{x})e^{\imath \phi (\tilde{x})}
\label{eqn:ampPhase}
\end{equation}
and obtain four boundary conditions.  From substituting Eq.~(\ref{eqn:ampPhase}) into Eq.~(\ref{eqn:pbcRealBC1}) and taking real and imaginary parts:

\begin{equation}
r(0) = r(1)
\label{eqn:pbcComplexBC1}
\end{equation}

\begin{equation}
\phi (1) - \phi (0) = 2\pi n
\label{eqn:pbcComplexBC2}
\end{equation}
where $n$ is an integer which we will call the phase quantum number.  From substituting Eq.~(\ref{eqn:ampPhase}) into Eq.~(\ref{eqn:pbcRealBC2}) and again taking real and imaginary parts:

\begin{equation}
[r^{\prime}\cos\phi - r \phi^{\prime}\sin\phi ]\mid_{\tilde{x}=0}=
[r^{\prime}\cos\phi - r \phi^{\prime}\sin\phi ]\mid_{\tilde{x}=1}
\label{eqn:pbcComplexBC3}
\end{equation}

\begin{equation}
[r^{\prime}\sin\phi - r \phi^{\prime}\cos\phi ]\mid_{\tilde{x}=0}=
[r^{\prime}\sin\phi - r \phi^{\prime}\cos\phi ]\mid_{\tilde{x}=1}
\label{eqn:pbcComplexBC4}
\end{equation}

Substituting Eq.~(\ref{eqn:ampPhase}) into Eq.~(\ref{eqn:nlse}), we divide the NLSE into real and imaginary parts.  We integrate once to solve for $\phi^{\prime}$ in the imaginary part, and find:

\begin{equation}
\phi^{\prime}=\frac{\alpha}{S}
\label{eqn:nlsePhase}
\end{equation}
where $\alpha$ is an undetermined constant of integration and $S\equiv r(\tilde{x})^{2}$ is the single particle density $\mid\!\!f(\tilde{x})\!\!\mid^{2}$.  Substituting Eq.~(\ref{eqn:nlsePhase}) into the real part, we multiply through by $r^{\prime}$ and integrate again.  We find:

\begin{equation}
(S^{\prime})^{2}=-2\:[\:-\frac{1}{\lambda^{2}}S^{3}+\frac{2\tilde{\mu}}{\lambda^{2}}S^{2}-\beta S+2\alpha^{2}\:]
\label{eqn:nlseDensity}
\end{equation}
where $\beta$ is an additional undetermined constant of integration.  A similiar solution method has been used by Drazin and Johnson in an elementary discussion of solitons\cite{drazin1}.  For the complex solutions, Eqs.~(\ref{eqn:nlsePhase})-(\ref{eqn:nlseDensity}) replace the NLSE as the equations to solve, together with boundary conditions (\ref{eqn:pbcComplexBC1})-(\ref{eqn:pbcComplexBC4}), and the normalization (\ref{eqn:normCond}).

We may rewrite Eq.~(\ref{eqn:nlseDensity}) as an integral:

\begin{equation}
\tilde{x}=\int_{0}^{S}\frac{dS}{\sqrt{2}\sqrt{(\lambda^{-2})S^{3}+(-2\tilde{\mu}\lambda^{-2})S^{2}+\beta S - 2\alpha^{2}}}
\label{eqn:nlseIntegral}
\end{equation}

This is an elliptic integral.  Any elliptic integral can be expressed as the sum of a finite number of elliptic integrals of the first, second, and third kinds.  Given that $\alpha$ and $\beta$ are real, such integrals may be reduced to a standard form with Cayley transformations, so that $0<m<1$ and the parameter $m$ is real\cite{bowman1}.  Therefore all intrinsically complex solutions to the 1D NLSE may be written as a sum over standard elliptic integrals.

We have found real symmetry-breaking solutions for which the density is proportional to sn$^{2}$.  These solutions vanish at $2j$ points around the ring.  We look for solutions of a similiar form for which the density does not vanish.  The physical motivation for such a solution type will become clear in Sec.~\ref{sec:connect}.  Using our physical intuition, we are able to bypass the use of Cayley transformations.

From the Jacobian elliptic identity Eq.~(\ref{eqn:jeIds}):

\begin{equation}
\text{sn}^{2}(\tilde{x}\mid m)=\frac{1}{m}(1-\text{dn}^{2}(\tilde{x}\mid m))
\end{equation}
where $\text{dn}(\tilde{x}\mid m)$ is the Jacobian elliptic function which we describe in Fig.~\ref{fig:Fjacobi}.  We thus generalize the real symmetry-breaking solutions, Eq.~(\ref{eqn:boxAnsatz}), as follows:

\begin{equation}
r^{2}(\tilde{x})=A^2(1+\gamma dn^{2}[k \tilde{x} + \delta \mid m))
\label{eqn:complexAnsatz0}
\end{equation}
where $-1\leq\gamma\leq 0$.

By setting $k$ equal to the full period of dn, i.e. $k=2jK(m)$, and $\delta=0$, we will automatically match the boundary conditions related to amplitude, Eqs.~(\ref{eqn:pbcComplexBC1}), (\ref{eqn:pbcComplexBC3}), and (\ref{eqn:pbcComplexBC4}).  It will remain to satisfy the phase quantization (\ref{eqn:pbcComplexBC2}).  For real solutions we said that $j$ was related to the number of nodes or kinks.  Here, as the density no longer goes to zero, $j$ is to be interpreted as the number of dips, or density-notches as we will call them, in the density $r(\tilde{x})^{2}$.  This is consistent with our previous definitions of $j$.  $\gamma$ is then the depth of the notch, while $A^2$ is put in to satisfy normalization.  We will consider the case of general $\delta$, and thus degeneracy, later.  

Eq.~(\ref{eqn:complexAnsatz0}) then becomes:

\begin{equation}
r^{2}(\tilde{x})=A^2(1+\gamma \text{dn}^{2}(2jK(m) \tilde{x}\mid m))
\label{eqn:complexAnsatz}
\end{equation}

Substituting this into Eq.~(\ref{eqn:nlseDensity}), using additional Jacobian elliptic identities, and setting coefficients of equal powers of dn equal, we obtain four equations in the parameters $\alpha$, $\beta$, $\gamma$, and $k$.  Eliminating $\beta$, we are left with $\alpha$, $\gamma$, and $A^2$ as a function of m, $\lambda$, $j$, and $\tilde{\mu}$.  We use the normalization (\ref{eqn:normCond}) to constrain $\tilde{\mu}$ and find:

\begin{equation}
\tilde{\mu}=\frac{3}{2}+12j^{2}\lambda^{2}E(m)K(m)-4j^{2}(2-m)\lambda^{2}K(m)^{2}
\label{eqn:complexEnergy}
\end{equation}
From this we obtain our equations for the parameters $\alpha$, $\gamma$, and $A^2$:

\begin{eqnarray}
\alpha & = & \frac{1}{\sqrt{2}}[\:(\frac{1}{\lambda})^{2}((1+ 8 j^2 \lambda^{2}E(m)K(m) \nonumber \\
       &   & - 8 j^{2} (1-m)\lambda^{2}K(m)^{2}) *( 1 - 8 j^2\lambda^2K(m)^{2} \nonumber \\
       &   & + 64 j^4 \lambda^{4} E(m)^{2} K(m)^{2} -16 j^2 \lambda^2 E(m) K(m) \nonumber \\
       &   & *(-1 + 4 j^{2}\lambda^2K(m)^{2}))) \:]^{1/2}
\label{eqn:alpha}
\end{eqnarray}

\begin{equation}
\gamma = - \frac{8 j^{2} \lambda^2 K(m)^{2}}{1+8 j^{2} \lambda^2 K(m) E(m)}
\end{equation}

\begin{equation}
A^2 = 1+8 j^{2} \lambda^2 K(m) E(m)
\end{equation}

This leaves the constant of integration $\alpha$ in $\phi^{\prime}=\alpha/r^{2}$, the depth $\gamma$, the prefactor to the density $A^2$, and the chemical potential $\tilde{\mu}$, determined up to the number of density-notches $j$, the scale $\lambda$, and the parameter $m$.  For a given $\lambda$ and $j$ we then numerically integrate the phase (\ref{eqn:nlsePhase}) and, using $m$ as our free parameter, adjust $m$ until the boundary condition (\ref{eqn:pbcComplexBC2}) is met, i.e. until the phase quantum number $n$ is an integer.  We note that all parameters are monotonic in $m$ so that our algorithm is quite straightforward.  By symmetry of the ring, $n$ can be either positive or negative, so that each solution is two-fold degenerate, just as we found for the constant amplitude solutions.

In Fig.~\ref{fig:FcomplexAmpPhase1} we show the amplitude and phase of one and two density-notch solutions at our test scale of $\lambda=1/25$.  We have plotted the amplitude above the phase to make apparent that the phase is a background constant slope with a region of increased slope where the density-notch occurs.  The deeper the notch, the larger the increase in slope.  In the limit that the notch dips to zero to form a node, the phase becomes a step function of height $\pi$ per step and the real solutions are recovered.

If $\delta$ is generalized so that it is arbitrary, a similiar degeneracy to what was found in Eq.~(\ref{eqn:entropy1}) results:

\begin{equation}
S\sim k_{\text{B}}ln(\frac{1}{\zeta j\lambda})
\label{eqn:entropy2}
\end{equation}
where $\zeta=\pi\sqrt{6}$, as we shall show in Sec.~\ref{subsec:pbcbounds}.

\subsection{Bounds}
\label{subsec:pbcbounds}

Real stationary states can have an arbitrarily large number of nodes.  But the number of notches for nodeless states is limited.  We set three bounds on the complex, nodeless solutions: the maximum chemical potential; the minimum and maximum phase quantum number; and the minimum scale to obtain $j$ notches.  As a consequence of these bounds there are some scales at which no complex solutions exist.

The maximum number of density-notches that can fit on the ring is obtained from the lower limit on the period of the dn function in Eq.~(\ref{eqn:complexAnsatz}).  When the notches overlap too much they are no longer solutions to the NLSE.  The dn function approaches its minimum period of $\pi$ as $m\rightarrow 0^{+}$.  In this limit Eq.~(\ref{eqn:complexEnergy}) is the maximum chemical potential:

\begin{equation}
\tilde{\mu}_{max}=\frac{1}{2}(3+2j^{2}\lambda^2\pi^{2})
\label{eqn:maxEnergy}
\end{equation}

In this same limit the amplitude approaches a constant which the normalization constrains to be 1.  From Eqs.~(\ref{eqn:nlsePhase}) and~(\ref{eqn:alpha}) we find a relation between the maximum number of density-notches, the phase quantum number, and the scale $\lambda$:

\begin{equation}
\lambda=\frac{1}{\pi \sqrt{8n^{2}-2j^{2}}}
\label{eqn:bounds}
\end{equation}

\begin{figure}
\begin{center}
\epsfig{figure=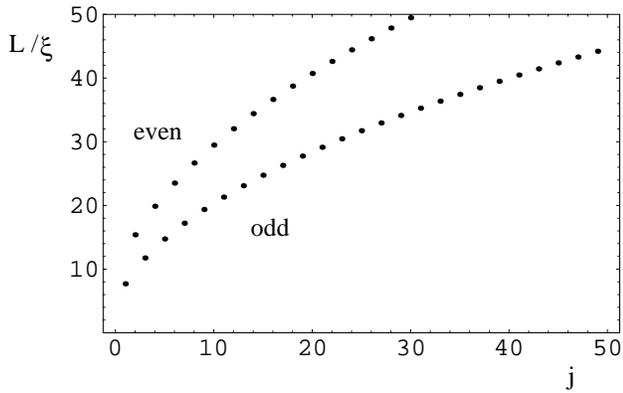,width=8.2cm}
\end{center}
\caption{
Minimum inverse scale for $j$ density-notches to become available.  The lower curve is odd $j$ and the upper curve is even $j$.  Note that at a given inverse scale there are many more odd solutions than even solutions available.  The ordering of the solutions is $j=(1,3,5,2,7,9,4,...)$.
}
\label{fig:FminScale}
\end{figure}

One may invert this relation to find the maximum $n$ for a given number of density-notches $j$.  Eq.~(\ref{eqn:bounds}) requires that $n>j/2$.  Thus there is both an upper and a lower bound on $n$:

\begin{equation}
\frac{j}{2} < n \leq \sqrt{\frac{1}{8}(\frac{1}{\pi^{2}\lambda^{2}}+2j^{2})}
\end{equation}
As $j$ increases $n_{max}\rightarrow(j/2)^{+}$.  Only integer $n$ can solve the phase quantization condition~(\ref{eqn:pbcComplexBC2}).  It follows that, for a given $\lambda^{-1}$, more odd $j$ solutions will be available than even $j$ solutions, because j/2 for odd $j$ is half integer.  This is apparent in Fig.~\ref{fig:FminScale}.

The above bounds imply that at a given scale the number of density notches is bounded from above.  In Fig.~\ref{fig:FminScale} we plot the scale at which $j$ density-notches become possible.  For less than 7.7 healing lengths to the box length, there are no complex solutions at all.  Then, in order, $j=1,3,5,2,7,9,4,...$ solutions become possible:

\begin{eqnarray}
\lambda^{-1}_{min}=\pi\sqrt{8+8j_{even}} \nonumber \\
\lambda^{-1}_{min}=\pi\sqrt{2+4j_{odd}}
\label{eqn:evenj}
\end{eqnarray}
for even and odd $j$, respectively


The minimum inverse scale for $j=1$ is $\lambda^{-1}=\pi\sqrt{6}$.  This is the natural size of a density-notch.  At smaller inverse scales the notches are affected by overlap, so that complex solutions do not exist, while real solutions become sinusoidal.  Thus the parameter $\zeta$ in Eq.~(\ref{eqn:entropy2}) takes the value $\zeta=\pi\sqrt{6}$.  Checking the limits, as $\lambda^{-1}\rightarrow (\lambda^{-1})_{min}=\pi\sqrt{6}$, $S\rightarrow 0^{+}$.  There is only one configuration.

\subsection{Spectra}
\label{subsec:pbcspectra}

\begin{figure}
\begin{center}
\epsfig{figure=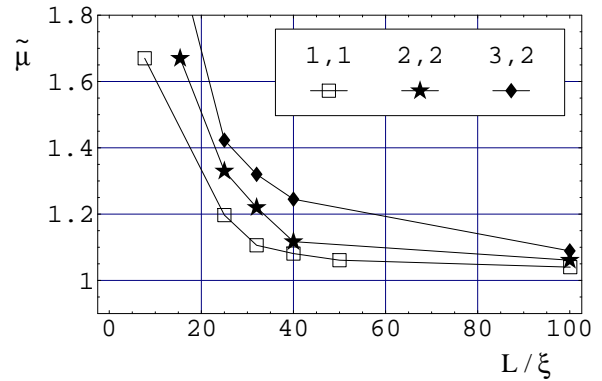,width=8.2cm}
\end{center}
\caption{
Chemical potential spectra for complex, symmetry-breaking stationary states on the ring as a function of inverse scale.  The three solutions shown here were found using the numerical algorithm prescribed in section \ref{subsec:pbccomplex}.  They are $(j,n)=(1,1),(2,2),(3,2),$ where $j$ is the number of density-notches and $n$ is the phase quantum number.  The leftmost points are the minimum inverse scale and maximum chemical potential possible for such a solution.  Bounds and ordering of the solutions are shown in Fig.~\ref{fig:FminScale}.
}
\label{fig:FcomplexSpectrum}
\end{figure}

We show the chemical potential spectra as a function of $\lambda^{-1}$ for the three types of stationary states on the ring: real, constant amplitude, and intrinsically complex.  In Fig.~\ref{fig:FbbcSpectrum} the two lowest real spectra are shown.  For comparison we have overlayed the four lowest constant amplitude spectra on the same figure.  In Fig.~\ref{fig:FcomplexSpectrum} we show the three lowest spectra for the complex solutions.

For our experimentally reasonable test scale of $\lambda=1/25$ the order of stationary states, starting with the ground state, is: constant amplitude, singly quantized vortex, single grey density notch, doubly quantized vortex, real two-node solution, two grey density notch, etc.  Note that the minimum chemical potential is $\tilde{\mu}=1$.

Since the real solutions are a limiting case of the complex solutions, the two scale in the same way and their energy levels do not cross.  But the constant amplitude solutions depend differently on inverse scale, so their energy levels can cross with those of the other solutions.


\section{Connection with soliton theory}
\label{sec:connect}

The dimensionless, time-dependent, free NLSE is:

\begin{equation}
\label{eqn:nlsetime}
[\imath\nu\partial_{t}+\xi^2\partial_{xx}-\mid\!f(x,t)\!\mid^{2}]f(x,t)=0
\end{equation}
where $\nu\equiv2M\xi^2/\hbar$ has units of time and we have chosen to use $x$ rather than $\tilde{x}$.  The single grey or dark density-notch solution to this equation takes the form\cite{kivshar3}:

\begin{equation}
f(x-ct,t)=\sqrt{2}\,[\imath\,\frac{c\nu}{2\xi}+\xi\kappa\,tanh(\kappa\,(x-c\,t))]\,e^{-\imath\mu t/\hbar}
\label{eqn:soliton}
\end{equation}
where $\kappa$ is the width, c is the speed, and $\mu$ is the chemical potential.

Using an identity to change from tanh to sech, the amplitude squared is:

\begin{equation}
\mid\!f(x-c\,t)\!\mid^{2}=\eta-2\xi^2\kappa^{2}sech^{2}(\kappa(x-c\,t))
\end{equation}
where $\eta\equiv((c^2\nu^2)/(2\xi^2)+2\kappa^2\xi^2)$.  The speed is:

\begin{equation}
c=\frac{\xi}{\nu}\sqrt{2}\sqrt{\eta-2\kappa^2\xi^2}
\label{eqn:speed}
\end{equation}
where c varies from zero to the Bogoliubov sound speed\cite{reinhardt1} which, on insertion of constants, is $c_{max}=\sqrt{4\pi\hbar^{2}a\bar{\rho}/m^{2}}$.  For $c=0$ we recover the dark, tanh soliton mentioned in Sec.~\ref{sec:bbc}.  For $c\rightarrow c_{max}^{-}$ the grey density notch approaches zero depth.  Thus the speed of the density notch has a finite range and the zero depth solitons approach the speed of sound from below.

\begin{figure}
\begin{center}
\epsfig{figure=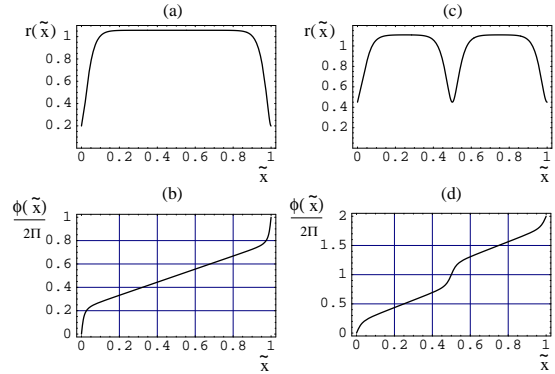,width=8.2cm}
\end{center}
\caption{
Complex, symmetry-breaking stationary solutions of the NLSE on a ring.  These are grey density-notch soliton trains.  A supercurrent around the ring exactly cancels their velocities to make them stationary states in the lab frame.  $j$ is the number of density-notches, $n$ is the phase quantum number, and $\tilde{\mu}$ is the chemical potential.  (a) Amplitude and (b) phase/$2\pi$ of the $j=1$, $n=1$, $\tilde{\mu}=1.197$ stationary state.  (c) Amplitude and (d) phase/$2\pi$ of the $j=2$, $n=2$, $\tilde{\mu}=1.331$ stationary state.  All plots are for the test scale of $\xi/L=1/25$.  
}
\label{fig:FcomplexAmpPhase1}
\end{figure}

The dn function is a periodic generalization of a sech.  Therefore replacing the sech in Eq.~(\ref{eqn:soliton}) by a dn is equivalent to replacing a soliton by a soliton train.  With this substitution, Eq.~(\ref{eqn:soliton}) becomes identical in form to Eq.~(\ref{eqn:complexAnsatz}) up to its time dependence.  A single soliton is set in motion by a phase jump~\cite{reinhardt1,kivshar3}.  The same holds true for a soliton train.  In the plots of the phase of the complex solutions, Figs.~\ref{fig:FcomplexAmpPhase1}b and~\ref{fig:FcomplexAmpPhase1}d, there was a background linear slope with a jump where the density-notches occurred.  Thus the complex stationary states can be interpreted as momentum boosts of the condensate with grey soliton-trains superimposed.  The velocity of the momentum boost exactly cancels the velocity of the soliton-train, resulting in a stationary state in the lab frame.

Although free grey soliton-trains can vary continuously in speed up to $c_{max}$, periodic boundary conditions require that the boost speed which brings about our stationary solutions is quantized.  Using Eq.~(\ref{eqn:speed}) and making the substitutions $\eta=A$ and $-2\kappa^{2}=A\gamma$ so that we change over to the constants used in Eq.~(\ref{eqn:complexAnsatz0}), the boost speed of the condensate which results in cancellation is:

\begin{equation}
c_{boost}=\frac{\xi}{\nu}\sqrt{2A(1+\gamma)}
\end{equation}
where $\gamma\rightarrow -1^{+}$ is the $c=0$ limit and $\gamma=0$ is the Bogoluibov sound speed.  $\gamma\rightarrow -1^{+}$ reproduces the real, dark soliton solutions found in Sec.~\ref{subsec:pbcreal}.  Thus the real solutions are soliton-trains at rest.

\section{Conclusion}
\label{sec:conclusion}

We have presented the complete set of stationary solutions to the nonlinear Schr\"odinger equation under periodic and box boundary conditions in one dimension for the case of repulsive nonlinearity.  In a box all solutions may be taken to be real.  On a ring there are three solution types: constant amplitude solutions which are plane waves; real symmetry-breaking solutions; and a novel class of complex symmetry-breaking solutions which correspond to a boost of the condensate in one direction with density-notches moving in the opposite direction, so that they are stationary in the lab frame.

Real and constant amplitude solutions are in one-to-one correspondence with those of the analogous particle-in-a-box and particle-on-a-ring problems in linear quantum mechanics.  Complex, symmetry-breaking solutions are uniquely nonlinear.  We showed that solutions of non-constant amplitude may be treated as density-notch soliton-trains.  As the natural size of a density-notch is $\pi\sqrt{6}$, the minimum scale size needed to obtain complex solutions is $L/\xi = \pi\sqrt{6}$.  In the context of the BEC, this means that the number of atoms determines the available solution-types.

The results of this paper cast the findings of previous work\cite{reinhardt1,reinhardt2} in the larger context of a comprehensive set of solutions to the NLSE.  In the previous work, perturbations of the phase of the purely real, box-type solutions were found to induce solitonic motion.  Such perturbations have recently been applied in the laboratory to three-dimensional \cite{denschlag1} and nearly quasi-one-dimensional\cite{burger1} BECs and have generated solitary waves.  Our present work gives a larger class of solutions, whose response to perturbations may suggest further experiments.


\acknowledgments

We benefited greatly from extensive discussions with Nathan Kutz and David Thouless.  Early stages of this work were supported in part by the Office of Naval Research; the work was completed with the partial support of NSF Chemistry and Physics.


\appendix
\section{Jacobian Elliptic functions}
\label{app:jacobian}

\begin{figure}
\begin{center}
\epsfig{figure=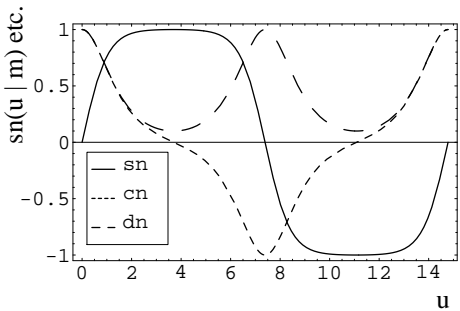,width=8.2cm}
\end{center}
\caption{
The three basic Jacobian elliptic functions, for the parameter $m=0.99$.  The solid line is sn, the dotted line is cn, and the dashed line is dn.  Note that the period of dn is half that of the other two and that dn is nodeless.  All Jacobian elliptic functions may be constructed from these three.  Of the twelve possible functions, these three shapes are the only normalizable ones not different from each other by a translation along the horizontal axis or a renormalization along the vertical axis.
}
\label{fig:Fjacobi}
\end{figure}

We here briefly review the properties of Jacobian elliptic functions and estabilsh the notation used herein.  There are a total of twelve such functions.  All of them solve the unbounded 1D NLSE in one form or another.  Of the twelve, six are normalizable.  Of these, only three have different physical form.  They are the sn, the cn, and the dn.  We plot them in Fig.~\ref{fig:Fjacobi}.  Of these, only the sn solves the NLSE for repulsive nonlinearity while only the cn and dn solve the NLSE for attractive nonlinearity.  The cd also solves the repulsive case.  It differs from the sn by a shift of a quarter period but is otherwise identical.

The function $\text{sn}(u\mid m)$ may be written in integral form:

\begin{equation}
u=\int_{0}^{x} \frac{dt}{\sqrt{1-t^{2}}\sqrt{1-m\,t^{2}}}
\label{eqn:snDefn}
\end{equation}
where $u=\text{sn}^{-1}(x)$ so that $x=\text{sn}(u\mid m)$.  The functions $\text{cn}(u\mid m)$ and $\text{dn}(u\mid m)$ may then be defined by the equations:

\begin{eqnarray}
\text{cn}(u\mid m)=\sqrt{1-\text{dn}^{2}(u\mid m)} \nonumber \\
\text{dn}(u\mid m)=\sqrt{1-m\,\text{sn}^{2}(u\mid m)}
\label{eqn:jeIds}
\end{eqnarray}

The limits of the sn, cn, and dn functions, along with the complete elliptic integrals $K(m)$ and $E(m)$ are listed in Table~\ref{table:jacobian}.  The period of the sn and cn functions is $4K(m)$, while that of the dn function is $2K(m)$.  We direct the reader to Abramowitz and Stegun\cite{abramowitz1} and other works\cite{bowman1} for a further review of the properties of these functions.


\begin{references}
\bibitem[*]{byline} to whom correspondence should be addressed
\end{references}

%
%

\begin{table}
\caption{Limits of Jacobian elliptic functions and integrals\cite{abramowitz1}.}
\label{table:jacobian}
\begin{tabular}{ccc}
  & $m=0$ & $m=1$ \\
\tableline
$\text{sn}(u \mid m)$ & $\sin (u)$  & $\text{tanh}(u)$ \\
$\text{cn}(u \mid m)$ & $\cos (u)$  & $\text{sech}(u)$ \\
$\text{dn}(u \mid m)$ & 1         & $\text{sech}(u)$ \\
$K(m)$                  & $\pi/2$   & $\infty$       \\
$E(m)$                  & $\pi/2$   & 1              \\
\end{tabular}
\end{table}


\end{document}